\documentclass[a4paper,12pt]{article}
\usepackage{epsfig,float}
\usepackage[left=2cm,right=2cm]{geometry}
\usepackage{a4wide,graphicx}
\newcommand\T{\rule{0pt}{4ex}}
\newcommand\B{\rule[-1.5ex]{0pt}{0pt}}
\newcommand{\eps}{\epsilon}

\begin{document}

\title{The $Y(3940)$, $Z(3930)$ and the $X(4160)$ as dynamically generated resonances from the vector-vector interaction}

\author{R. Molina$^1$ and E. Oset$^1$}
\maketitle

\begin{center}
$^1$ Departamento de F\'{\i}sica Te\'orica and IFIC,
Centro Mixto Universidad de Valencia-CSIC,
Institutos de Investigaci\'on de Paterna, Aptdo. 22085, 46071 Valencia,
 Spain\\
\end{center}

\date{}

 \begin{abstract}  

We study the vector-vector interaction within the framework of the hidden gauge formalism for the channels with quantum numbers Charm $C=0$ and Strangeness $S=0$ in the energy region around $4000$ MeV. By looking for poles in the complex plane we find three resonances that could be identified by the mass, width and quantum numbers with the $Y(3940)$, $Z(3940)$ and $X(4160)$, these poles appear with isospin $I=0$ and $J^{PC}=0^{++}$, $2^{++}$ and $2^{++}$ respectively. Whereas the $Y(3940)$ and  $Z(3940)$ are coupled more strongly to $D^*\bar{D}^*$, the $X(4160)$ is basically a $D^*_s \bar{D}^*_s$ molecular state. Another two extra resonances appear in our approach with $I=0, 1$ and $J^{PC}=1^{+-}, 2^{++}$ which are not found in the PDG with masses $M=3945, 3912$ MeV and widths $\Gamma=0, 120$ MeV respectively. 
\end{abstract}

\section{Introduction}

     The charmonium spectroscopy has been recently pushed forward with the unexpected discovery of many new charmonium-like resonances in the $B$ factories at SLAC, KEK and CESR. The $B$-factories, originally constructed to test matter-antimatter asymmetries or CP-violation, within or beyond the standard model, discovered a number of interesting charm and hidden charm mesons which do not seem to have a simple $c\bar{c}$ structure. The first of these XYZ states is the $X(3872)$ that was observed by the Belle Collaboration as a narrow peak near $3872$ MeV in the $\pi^+\pi^-J/\psi$ invariant mass distributions in $B^-\to K^-\pi^+\pi^-J/\psi$ decays \cite{Choi,Acosta,Abazov,Aubert1}. Belle also observed that the rate of the decay $X(3872)\to \pi^+\pi^-\pi^0 J/\psi$ was comparable to $\pi^+\pi^- J/\psi$ \cite{Abe}, with the $3\pi$ or $2\pi$ coming from $\omega$ or $\rho$ which would imply the $X(3872)$ is a mixture of both $I=0$ and $I=1$ \cite{Close} (see however \cite{Swanson,Braaten,Gamermann:2009fv} for alternative explanations of that ratio). The close value of the $X(3872)$ mass to the sum of the masses $m_{D^0}+m_{D^{*0}}=3871.81\pm 0.36$ MeV led to consider the $X(3872)$ as a molecule-like bound state of a $D^0$ and a $\bar{D}^{*0}$ meson, and much work on the properties of these possible systems has been done \cite{Close,Swanson,Voloshin,De Rujula,Tornqvist,Voloshin2,Gamermann:2007fi}.
     
     The next XYZ states, that we will consider in this paper, are the $X(3940)$, the $Y(3940)$, the $X(4160)$ and the $Z(3930)$. The $X(3940)$ was observed in the double-charmonium production reaction $e^+e^-\to J/\psi+X$ with mass $M=3943\pm8$ MeV and width $\Gamma<52$ MeV \cite{Abe2}. After that Belle also observed a $D^*\bar{D}^*$ mass peak in the $e^+e^-\to J/\psi D^*\bar{D}^*$ reaction \cite{Abe3}. Whereas the $X(3940)\to D\bar{D}^*$ has been observed, there is no signal for the $D\bar{D}$ or the $\omega J/\psi$ decays. Because of that and the fact that the $\eta_c(1S)$ and $\eta_c(2S)$ were also produced in double-charm production, it was believed that the $X(3940)$ could have $J^{PC}=0^{-+}$, being a $3^1 S_1$ charmonium state ($\eta''_c$), but there are problems with this assignment since in this case the $X(3940)$ should have a mass $\sim$ $4050$ MeV or even higher \cite{Barnes}. Thus, it seems very unlikely that the $X(3940)$ is a $c\bar{c}$ state. 
     
     Belle has observed a state with a decay mode $Y(3940)\to \omega J/\psi$ in $B\to K\omega J/\psi$ decays \cite{Abe4} and Babar has confirmed it \cite{Aubert}, although the values for the mass and width reported by Babar are smaller than Belle's values ($M=3943\pm17$ MeV and $\Gamma=87\pm 34$ MeV is reported by Belle and $M=3914.3^{+4.1}_{-3.8}$ MeV and $\Gamma=33^{+12}_{-8}$ MeV by Babar). In principle, the mass and width of the $Y(3940)$ suggest a radially excited P-wave charmonium state but then the $\chi_{c1}(2P)\to D \bar{D}^*$ would be the dominant decay mode and has not been observed. Hence, it seems very unlikely that the reported $Y(3940)$ and $X(3940)$ represent different decay modes of the same state \cite{Olsenmult}. The $J^{PC}$ assignment in the case that the $Y(3940)$ were a charmonium state is not very clear. Indeed, for a charmonium state with $J^{PC}=0^{-+}$ ($\eta_c$) the mass is a little low and for $J^{PC}=0^{++} (\chi'_{c0})$ the mass is too high \cite{Olsenmult}, also if it had a simple $c\bar{c}$ structure one would expect that the open charm decay modes would be dominant, and the other,  $\omega J/\psi$, negligible. Also, the large partial width $Y(3940)\to \omega J/\psi$, estimated above $1$ MeV \cite{Godfrey}, is quite larger than the measured partial widths for any of the observed hadronic transitions between charmonium states.
     
     Belle has recently observed a mass peak in the $D^*\bar{D}^*$ system in the $e^+e^-\to J/\psi D^*\bar{D^*}$ reaction \cite{Abe3}. This state with a mass of $(4156\pm 29)$ MeV and a width of $\Gamma=139^{+113}_{-65}$ MeV has been called $X(4160)$. The production mechanism ensures that it has $C=+$. The known charmonium states seen from a $e^+e^-\to J/\psi D^*\bar{D^*}$ reaction have $J=0$, thus, although this state could be identified with a $3^1S_0$ ($\eta''_c$) or $4^1S_0$ ($\eta'''_c$) charmonium state, the mass predicted in the first case would be smaller ($4050$ MeV) and higher ($4400$ MeV) in the second case \cite{Barnes}. Recently, the CDF Collaboration at Fermilab has announced a narrow peak near the $J/\psi \phi$ threshold, which is designated as $Y(4140)$, observed in the $B^+\to J/\psi \phi K^+$ decay, it has a mass $M=4143\pm 2.9 (stat) \pm 1.2 (syst)$ MeV and a width $\Gamma=11.7^{+8.3}_{-5.0} (stat) \pm 3.7 (syst)$ MeV \cite{CDF}. The width observed is quite different from the one reported by Belle for the $X(4160)$ \cite{Abe3}, suggesting that one is taking about a different state. Another point is the structure and quantum numbers. There are some predictions: In \cite{Liu} the authors solved the Schr\"odinger equation from the potential obtained using effective Lagrangians and they found molecular solutions for $Y(3930)$ and $Y(4140)$ with $J^P=0^+, 2^+$, concluding that the $Y(3930)$ and the $Y(4140)$ are molecular partners. In \cite{Tania}, the authors assume that the $Y(3940)$ and the $Y(4140)$ are hadronic molecules with quantum numbers $J^{PC}=0^{++}$ or $2^{++}$ whose constituents are the charm vectors $D^*\bar{D}^*$ for the $Y(3940)$ and $D_s^{*+}\bar{D}_s^{*-}$ for the $Y(4140)$ and they calculate the decay rates of the observed modes $Y(3940)\to J/\psi \omega$ and $Y(4140)\to J/\psi$ for the case $J^{PC}=0^{++}$. The coupling constants are determined by means of the compositeness condition \cite{Tania} and the results for these decay modes support the molecular interpretation of the $Y(3930)$ and the $Y(4140)$. In \cite{Wang,Bracco} the authors use QCD sum rules to evaluate the mass of a possible mesonic state that couples to a molecular $D_s^{*+}\bar{D}_s^{*-}$ current. Whereas in \cite{Bracco} a mass $M=4.14\pm 0.09$ GeV is found, concluding that it is possible to describe the $Y(4140)$ as a $D_s^* \bar{D}_s^*$ molecular state, in \cite{Wang} they find a larger value of the mass $M=4.43\pm 0.16$ GeV. In \cite{Chao} the authors suggest possible assignments for the charmonium-like "X, Y, Z" states, for example the $X(4160)$ could be assigned to a $\chi_{c0}(3P)$ state. Also in \cite{Liu2} the authors find that the P-charmonium $\chi''_{cJ} (J=0,1)$ associated to the $Y(4140)$ is problematic because of the upper limit of the branching ratio of the decay $Y(4140)\to J/\psi \phi$ computed of the order of $10^{-4}-10^{-3}$. In \cite{stancu} a tetraquark structure is suggested for this state, while in \cite{beve} it is argued than the peak at this energy is just a $\phi J/\psi$ threshold effect.
     
     The $Z(3930)$ is also reported by Belle as a peak in the spectrum of $D\bar{D}$ mesons produced in $\gamma \gamma$ collisions, with mass and width $M=3929\pm 6$ MeV and $\Gamma =29\pm 10$ MeV. In the production process the two photons can only produce $D\bar{D}$  in a $0^{++}$ or $2^{++}$ states. The Belle measurements favors the $2^{++}$ hypothesis, making the assignment of the $Z(3930)$ to the $2^3 P_2 (\chi'_{c2})$ charmonium state possible \cite{Godfrey}, following the arguments given in \cite{Chao}.
     
     In this paper we will propose a theoretical explanation on the nature of some of these XYZ states, providing structure and quantum numbers for them. Our work is based on the hidden gauge symmetry (HGS) formalism for the interaction of vector mesons, which was introduced by Bando-Kugo-Yamawaki \cite{hidden1,hidden2,hidden3}. This hidden gauge formalism has been applied in \cite{raquel} for the $\rho \rho$ interaction, giving rise to two bound $\rho\rho$ states that could be identified with the $f_0(1370)$ and $f_2(1270)$. After that, the radiative decay of these states into $\gamma \gamma$ was studied in \cite{junko}, obtaining results in good agreement with the PDG \cite{pdg}. The work of \cite{raquel} was later extended to $SU(3)$ \cite{geng}, where several states were found that could also be identified with some of the PDG. Recently, the $\rho D^*$, $\omega D^*$ interaction was studied within the same formalism \cite{raquel2}, generalizing to $SU(4)$ the basic hidden gauge Lagrangians, but breaking the $SU(4)$ symmetry in the vector exchange diagrams. Some states were found which could be identified with the $D^*_2(2460)$ and the $D^*(2640)$, giving a prediction of the quantum numbers for the last one $I(J^P)=1/2(1^+)$ in the case the correspondence was correct, and providing a reasonable explanation for the small width of this state. Also, a new $D_0$ state was predicted with $J^P=0^+$, a mass close to $2600$ MeV and width of about $60$ MeV.
     
     The present paper follows the steps of \cite{raquel2}, investigating possible vector - vector states of hidden charm, using a unitary approach in coupled channels including all the channels involved for Charm $C=0$ and Strangeness $S=0$.

\section{Formalism for $VV$ interaction}
\subsection{Lagrangian}

\begin{figure}
\begin{center}
\includegraphics[width=16cm]{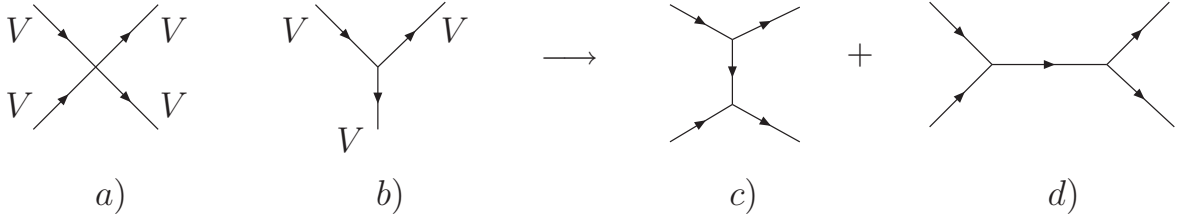}
\end{center}
\caption{Terms of the ${\cal L}_{III}$ Lagrangian: a) four vector contact term,
 Eq.~(\ref{lcont}); b) three-vector interaction, Eq.~(\ref{l3V}); c) $t$ and 
 $u$ channels from vector exchange; d) $s$ channel for vector exchange.}
\label{fig:fig1} 
\end{figure}

Our starting point is the Lagrangian, which involves the interaction of 
vector mesons amongst themselves, coming from the formalism of the hidden gauge symmetry (HGS) for vector mesons \cite{hidden1,hidden2,hidekoroca}
\begin{equation}
{\cal L}_{III}=-\frac{1}{4}\langle V_{\mu \nu}V^{\mu\nu}\rangle \ ,
\label{lVV}
\end{equation}
where the symbol $\langle \rangle$ stands for the trace in the $SU(4)$ space 
and $V_{\mu\nu}$ is given by 
\begin{equation}
V_{\mu\nu}=\partial_{\mu} V_\nu -\partial_\nu V_\mu -ig[V_\mu,V_\nu]\ ,
\label{Vmunu}
\end{equation}
with $g$ given by
\begin{equation}
g=\frac{M_V}{2f}\ ,
\label{g}
\end{equation}
and $f=93$ MeV the pion decay constant. Using the value of $g$ in Eq.~(\ref{g}) is 
one of the ways to account for the KSFR relation \cite{KSFR} which
is tied to  
vector meson dominance \cite{sakurai}.
The vector field $V_\mu$ is represented by the $SU(4)$ matrix which is
parametrized by 16 vector mesons including the 15-plet and singlet of $SU(4)$,
\begin{equation}
V_\mu=\left(
\begin{array}{cccc}
\frac{\rho^0}{\sqrt{2}}+\frac{\omega}{\sqrt{2}}&\rho^+& K^{*+}&\bar{D}^{*0}\\
\rho^-& -\frac{\rho^0}{\sqrt{2}}+\frac{\omega}{\sqrt{2}}&K^{*0}&D^{*-}\\
K^{*-}& \bar{K}^{*0}&\phi&D^{*-}_s\\
D^{*0}&D^{*+}&D^{*+}_s&J/\psi\\
\end{array}
\right)_\mu \ ,
\label{Vmu}
\end{equation}
where the ideal mixing has been taken for $\omega$, $\phi$ and $J/\psi$.
The interaction of ${\cal L}_{III}$ gives rise to a contact term coming from 
$[V_\mu,V_\nu][V_\mu,V_\nu]$
\begin{equation}
{\cal L}^{(c)}_{III}=\frac{g^2}{2}\langle V_\mu V_\nu V^\mu V^\nu-V_\nu V_\mu
V^\mu V^\nu\rangle\ ,
\label{lcont}
\end{equation}
depicted in Fig.~\ref{fig:fig1} a), and on the other hand it gives rise to a three 
vector vertex
\begin{equation}
{\cal L}^{(3V)}_{III}=ig\langle (\partial_\mu V_\nu -\partial_\nu V_\mu) V^\mu V^\nu\rangle
\label{l3V}\ ,
\end{equation}
depicted in Fig.~\ref{fig:fig1} b). This latter Lagrangian gives rise to a
$VV\to VV$ interaction by means of the exchange of one of the vectors, as 
shown in Figs.~\ref{fig:fig1} c), d).

The $SU(4)$ structure of the Lagrangian allows us to take into account all the
channels within $SU(4)$ which couple to certain quantum numbers. 
In the present work we shall 
present results for the case of all the channels which couple to $D^*\bar{D}^*$ or $D_s^*\bar{D}_s^*$. The formalism is
the same used in \cite{raquel} and \cite{raquel2}. Some approximations were made there which make
the formalism handy and reliable, by neglecting the three-momentum of the vector
mesons with respect to their masses. It is interesting to see that with this
approximation one obtains \cite{hidekoroca} from the hidden gauge approach 
the chiral local
Lagrangians which are used to study the interaction of pseudoscalar mesons
among themselves and the pseudoscalar mesons with vector mesons and with baryons
\cite{Gasser:1984gg,ulf,Birse:1996hd,Ecker:1989yg}.

\subsection{Four-vector contact interaction}

The channels that we are interested in are those with Charm $C=0$ and Strangeness $S=0$, they are, in the case of $I=0$: 
\begin{center}
$\mathbf{D^*\bar{D}^*}(4017)$, $\mathbf{D^*_s\bar{D}^*_s}(4225)$, $\mathbf{K^*\bar{K}^*}(1783)$, $\mathbf{\rho\rho}(1551)$, $\mathbf{\omega\omega}(1565)$\\\vspace{0.5cm}  $\mathbf{\phi\phi}(2039)$, $\mathbf{J/\psi J/\psi} (6194)$, $\mathbf{\omega J/\psi} (3880)$, $\mathbf{\phi J/\psi} (4116)$, $\mathbf{\omega \phi} (1802)$
\end{center}
where the magnitude between parenthesis is the sum of the masses of the two meson involved, and for $I=1$:
\begin{center}
$\mathbf{D^*\bar{D}^*}(4017)$, $\mathbf{K^*\bar{K}^*}(1783)$, $\mathbf{\rho\rho}(1551)$, $\mathbf{\rho\omega}(1558)$, $\mathbf{\rho J\psi}(3872)$, $\mathbf{\rho\phi} (1795)$.
\end{center}
We are not interested in the case of $I=2$ which was considered in \cite{geng} and where no bound states or resonances were found. 

\begin{figure}
\begin{center}
\includegraphics{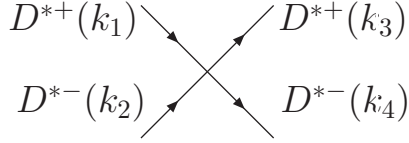}
\end{center}
\caption{Contact term of the $D^{*+}D^{*-}$ interaction.}
\label{fig:fig2}
\end{figure}

Consider now the  $D^{*+} D^{*-}\to D^{*+} D^{*-}$ reaction, see Fig.~\ref{fig:fig2}. In order to get the amplitude we use the Lagrangian of Eq.~(\ref{lcont}) obtaining
\begin{eqnarray}
t^{(c)}_{D^{*+}D^{*-}\to D^{*+}D^{*-}}= 2\,g^2 (\eps^{(1)}_\mu \eps^{(2)}_\mu \eps^{(3)\,\nu} \eps^{(4)\,\nu} +\eps^{(1)}_\mu \eps^{(2)}_\nu \eps^{(3)\,\mu} \eps^{(4)\,\nu} -2\eps^{(1)}_\mu \eps^{(2)}_\nu \eps^{(3)\,\nu} \eps^{(4)\,\mu} )\ ,
\label{eq:rhoDvertex}
\end{eqnarray}
where the indices $1,2,3$ and $4$ correspond to the particles with momenta 
$k_1,k_2,k_3$ and $k_4$ in Fig.~\ref{fig:fig2}. 
It is straightforward to write down all amplitudes for the other channels.

 In the approximation of neglecting the three-momenta of the vector mesons, only
 the spatial components of the polarization vectors are nonvanishing, the orbital angular momentum is $L=0$, and then
 one can obtain easily spin projection operators \cite{raquel} into spin
0, 1, 2 states,
  which are given below:
  
  \begin{eqnarray}
{\cal P}^{(0)}&=& \frac{1}{3}\eps_\mu \eps^\mu \eps_\nu \eps^\nu\nonumber\\
{\cal P}^{(1)}&=&\frac{1}{2}(\eps_\mu\eps_\nu\eps^\mu\eps^\nu-\eps_\mu\eps_\nu\eps^\nu\eps^\mu)\nonumber\\
{\cal P}^{(2)}&=&\lbrace\frac{1}{2}(\eps_\mu\eps_\nu\eps^\mu\eps^\nu+\eps_\mu\eps_\nu\eps^\nu\eps^\mu)-\frac{1}{3}\eps_\mu\eps^\mu\epsilon_\nu\epsilon^\nu\rbrace\ ,
\label{eq:projmu}
\end{eqnarray}
  where the order $1,\,2,\,3,\,4$ in the polarization vectors is
understood. We can then write the combination of polarization vectors
appearing in Eq.~(\ref{eq:rhoDvertex}) in terms of the spin combinations
and thus we obtain the kernel of the 
   interaction which will be later on used to solve the Bethe-Salpeter
   equation. However, it is practical to construct the isospin combinations
   before the spin projection is done.

     Recalling that we have isospin doublets $(-D^{*0},D^{*+})$ and $(D^{*-},\bar{D}^{*0})$, the $I=0$ and $1$ combinations are written as 
     
      \begin{eqnarray}
      |D^* \bar{D}^*,I=0,I_3=0\rangle&=&\frac{1}{\sqrt{{2}}}|D^{*+} D^{*-}\rangle+\frac{1}{\sqrt{2}}|D^{*0} \bar{D}^{*0}\rangle,\nonumber\\
      |D^* \bar{D}^*,I=1,I_3=0\rangle&=&\frac{1}{\sqrt{{2}}}|D^{*+} D^{*-}\rangle-\frac{1}{\sqrt{2}}|D^{*0} \bar{D}^{*0}\rangle .
      \label{eq:isospincomb}
      \end{eqnarray}
We then find the amplitudes in the isospin base by forming linear
combinations of the amplitudes in the particle base weighted by the
Clebsh-Gordan coefficients as given in Eq.~(\ref{eq:isospincomb}),
       \begin{eqnarray}
t^{(I=0)}&=&3\,g^2(\eps_\mu\eps_\mu\eps^\nu\eps^\nu+\eps_\mu\eps^\nu\eps_\mu\eps^\nu-2\eps_\mu\eps_\nu\eps^\nu\eps^\mu)\ ,\nonumber\\
t^{(I=1)}&=&g^2(\eps_\mu\eps_\mu\eps^\nu\eps^\nu+\eps_\mu\eps_\nu\eps^\mu\eps^\nu-2\eps_\mu\eps^\nu\eps_\nu\eps^\mu)\ .
\label{eq:rhoDIsospin}
\end{eqnarray}
These amplitudes, after
      projection into the spin channels, give rise to the following kernels
      (potential) for $I=0$,
      \begin{eqnarray}
      t^{(I=0,J=0)}&=&+6 g^2\ ,\nonumber\\
      t^{(I=0,J=1)}&=&+9 g^2\ ,\nonumber\\
      t^{(I=0,J=2)}&=&-3 g^2\ ,
      \label{eq:rhoDIS}
      \end{eqnarray}
and also for the case of $I=1$,
\begin{eqnarray}
      t^{(I=1,J=0)}&=&2 g^2\ ,\nonumber\\
      t^{(I=1,J=1)}&=&3 g^2\ ,\nonumber\\
      t^{(I=1,J=2)}&=&- g^2\ .
      \label{eq:rhoDIS3}
      \end{eqnarray}
      
      In the same way, we proceed to calculate the contact term projected in isospin and also spin between two of anyone of the involved channels.

\subsection{Vector meson exchange terms}

\begin{figure}
\begin{center}
\includegraphics{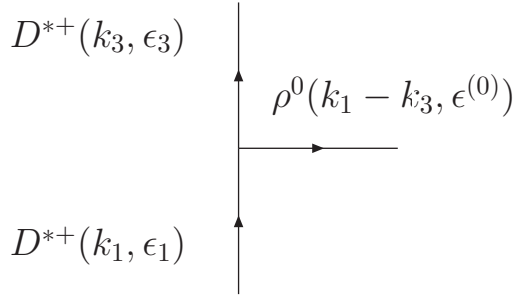}
\end{center}
\caption{three-vector vertex diagram.}
\label{fig:fig5}
\end{figure}

\begin{figure}
\begin{center}
\includegraphics{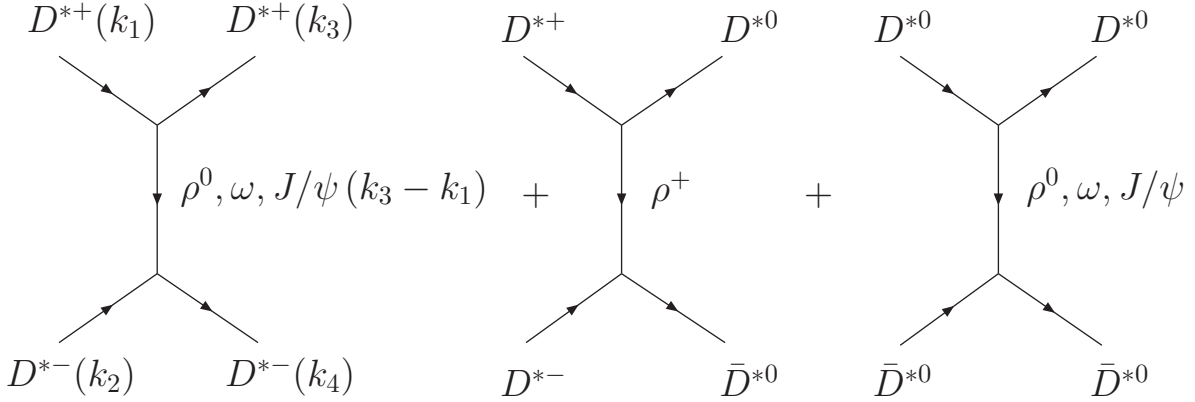}
\end{center}
\caption{Vector exchange diagrams for $D^{*} \bar{D}^{*}\to D^{*} \bar{D}^{*}$.}
\label{fig:fig6}
\end{figure}

Continuing with the $D^{*+}D^{*-}\to D^{*+}D^{*-}$ reaction, now we want to calculate the amplitude of the first diagram in Fig.~\ref{fig:fig6}. In order to do that, we follow the same procedure as in \cite{raquel2}, starting off the three-vector vertex which is given by the Lagrangian of Eq.~(\ref{l3V}) and is depicted in Fig.~\ref{fig:fig5} in the case of $D^{*+}\to D^{*+} \rho^0$.
  The amplitude of Fig.~\ref{fig:fig5} can be automatically calculated rewriting the Lagrangian
  of Eq.~(\ref{l3V}) as,
  \begin{eqnarray}
{\cal L}^{(3V)}_{III}=ig\langle V^\nu\partial_\mu V_\nu V^\mu-\partial_\nu V_\mu
V^\mu V^\nu\rangle \nonumber\\
=ig\langle (V^\mu\partial_\nu V_\mu -\partial_\nu V_\mu
V^\mu) V^\nu\rangle
\label{l3Vsimp}\ .
\end{eqnarray}
As it was explained in \cite{raquel2}, within the approximation of neglecting the three-momenta of external vectors, the  
 $V^\nu$ corresponds to the exchanged vector, simplifying the
 calculation. As already mentioned, this corresponds to the consideration of only the $s$-wave.

The vertex function corresponding to the diagram of Fig.~\ref{fig:fig5} is given 
by
\begin{eqnarray}
t^{(3)}=\frac{g}{\sqrt{2}}\, (k_{1} + k_{3})_\mu \eps_{1\,\nu}\eps^{\nu}_3 \eps^{(0)}_\mu
\label{vertexfig3}
\end{eqnarray}
With this basic structure,  and considering all the particles involved in the exchange, we can readily evaluate the amplitude of the first diagram 
of Fig.~\ref{fig:fig6} to obtain 

\begin{eqnarray}
t^{(ex)}_{D^{*+} D^{*-}\to D^{*+} D^{*-}}=-\frac{1}{2} \,g^2 (\frac{2}{M^2_{J/\psi}}+\frac{1}{M^2_\rho}+\frac{1}{M^2_\omega}) (k_1+k_3)\cdot(k_2+k_4)\, \eps_{1\,\mu}\eps_{2\,\nu}\eps_{3}^\mu\eps_{4}^\nu
\label{exchange}
\end{eqnarray}

In the vector exchange diagrams the exchange of one heavy-vector meson becomes supressed with respect to the exchange of a light vector by the ratio $M^2_L/M^2_H$ with $L=\omega, \rho$ and $H=J/\psi$. To project in s-wave one must do the following replacements:
\begin{eqnarray}
k_1\cdot k_2 &=& \frac{s-M^2_1-M^2_2}{2}\nonumber\\
k_1\cdot k_3 &=&k^0_1 k^0_3-\vec{p}\cdot \vec{q}\to \frac{(s+M^2_1-M^2_2) (s+M^2_3-M^2_4)}{4 s}\nonumber
\end{eqnarray}
where $\to$ means the projection over s-wave, and $k_1=(k^0_1,\vec{p})$, $k_2=(k^0_2,-\vec{p})$,  $k_3=(k^0_2,\vec{q})$, $k_4=(k^0_4,-\vec{q})$ and $M_i$, with $i=1,4$, is the mass of each particle, and so on for the other products of momenta. Thus, Eq. (\ref{exchange}) can be written as 
\begin{equation}
t^{(ex)}_{D^{*+} D^{*-}\to D^{*+} D^{*-}}=\frac{g^2 (2\,M^2_\omega M^2_\rho+M^2_{J/\psi} (M^2_\omega+M^2_\rho)) (4\,M^2_{D^*}-3\, s) \eps_{1\,\mu}\eps_{2\,\nu}\eps_3^\mu \eps_4^\nu}{4\, M^2_{J/\psi} M^2_\omega M^2_\rho}
\end{equation}

 The isospin projections taking into account the three diagrams of Fig. \ref{fig:fig6} and Eq. (\ref{eq:isospincomb}) give us
 \begin{eqnarray}
 t^{(ex,I=0)}_{D^{*} \bar{D}^{*}\to  D^{*}\bar{D}^{*}}&=&\frac{g^2 (2\,M^2_\omega M^2_\rho+M^2_{J/\psi} (3\,M^2_\omega+M^2_\rho)) (4\,M^2_{D^*}-3\, s) \eps_{1\,\mu}\eps_{2\,\nu}\eps_3^\mu \eps_4^\nu}{4\, M^2_{J/\psi} M^2_\omega M^2_\rho}\nonumber\\
 t^{(ex,I=1)}_{D^{*} \bar{D}^{*}\to  D^{*}\bar{D}^{*}}&=&\frac{g^2 (2\,M^2_\omega M^2_\rho+M^2_{J/\psi} (-M^2_\omega+M^2_\rho)) (4\,M^2_{D^*}-3\, s) \eps_{1\,\mu}\eps_{2\,\nu}\eps_3^\mu \eps_4^\nu}{4\, M^2_{J/\psi} M^2_\omega M^2_\rho}\ .
 \label{eq:exchisospin}
 \end{eqnarray}
Now using the equations for the spin projections we can split the terms into
 their spin parts and we obtain
 \begin{eqnarray}
 t^{(ex,I=0,S=0,1,2)}_{D^* \bar{D}^{*}\to D^* \bar{D}^{*}}&=&\frac{g^2 (2\,M^2_\omega M^2_\rho+M^2_{J/\psi} (3\,M^2_\omega+M^2_\rho)) (4\,M^2_{D^*}-3\, s) }{4\, M^2_{J/\psi} M^2_\omega M^2_\rho}\nonumber\\
 t^{(ex,I=1,S=0,1,2)}_{D^* \bar{D}^{*}\to D^* \bar{D}^{*}}&=&\frac{g^2 (2\,M^2_\omega M^2_\rho+M^2_{J/\psi} (-M^2_\omega+M^2_\rho)) (4\,M^2_{D^*}-3\, s) }{4\, M^2_{J/\psi} M^2_\omega M^2_\rho}\ .
 \label{eq:exchisospinspin}
 \end{eqnarray}
 For the vector exchange diagrams we obtain the same expression for all the spin states. One must do the same calculations for every channel, but, one can see that all the terms corresponding to the t-channel have the structure 
 \begin{center}
 $(k_1+k_3)\cdot (k_2+k_4) \eps_{1\,\mu} \eps_{2\, \nu}\eps^\mu_3 \eps^{\nu}_4$
 \end{center}
 and those that corresponds to the u-channel 
  \begin{center}
 $(k_1+k_4)\cdot (k_2+k_3) \eps_{1\,\mu} \eps_{2\, \nu}\eps^{\nu}_3\eps^\mu_4$
 \end{center}
 
 In Tables \ref{tab:Am1}, \ref{tab:Am2}, \ref{tab:Am3}, \ref{tab:Am4}, \ref{tab:Am5} and  \ref{tab:Am6} of the Appendix we give the $D^*\bar{D}^*\to channel$ and $D_s^* \bar{D}_s^*\to channel$ amplitudes projected into isospin and spin. The amplitudes, $J/\psi J/\psi$, $\omega J/\psi$, $\phi J/\psi\to K^*\bar{K}^*$, $\rho\rho$, $\omega\omega$, $\phi\phi$, $J/\psi J/\psi$, $\omega J/\psi$, $\phi J\psi$, $\omega\phi$ for $I=0$ and $\rho J/\psi \to K^* \bar{K}^*$, $\rho\rho$, $\rho\omega$, $\rho J/\psi$, $\rho\phi$ for $I=1$, are not in the tables because the interaction is exactly zero in our model. Also, the $SU(3)$ amplitudes that involve the $K^*\bar{K}^*$, $\rho\rho$, $\omega\omega$, $\phi\phi$ and $\omega\phi$ channels are in the Tables V-X (contact term) and XVIII-XXI (exchange term) in \cite{geng}. Here we have used the constant $g=M_\rho/(2 f_\pi)=4.17$ with $f_\pi=93$ MeV for light mesons and $g_D=M_{D^*}/(2 f_{D})=6.9$, $g_{D_s}=M_{D^*_s}/(2 f_{D_s})=5.47$, $g_{\eta_c}= M_{J/\psi}/(2 f_{\eta_c})=5.2$ with $f_D=206/\sqrt{2}=145.66$ MeV \cite{pdg}, $f_{D_s}=273/\sqrt{2}=193.04$ MeV \cite{pdg}, $f_{\eta_c}=420/\sqrt{2}$ MeV \cite{Sun} for the $D^*$, $D^*_s$ and $J/\psi$ mesons\footnote{We have made the approximation $\sqrt{g g_{\eta_c}}\sim g$ in the channels that involve one light meson and the $J/\psi$ meson.}. As we see in these tables the exchange term is equal for $J=0$ and $J=2$ and the exchange of one heavy vector meson is supressed by the factor $\kappa=M^2_L/M^2_{H}$ which is equal to $0.15$ in the case of $M_L=M_\rho$ and $M_H=M_{D^*}$. We can observe that the interaction at the $D^*\bar{D}^*$ threshold becomes attractive and very strong for $D^*\bar{D}^*\to D^*\bar{D}^*$, $D^*\bar{D}^*\to D_s^*\bar{D}_s^*$ and $D_s^*\bar{D}_s^*\to D_s^*\bar{D}_s^*$ amplitudes, which will lead to bound states of two vector mesons, basically $D^*\bar{D}^*$ and/or $D^*_s\bar{D}^*_s$, in the region of interest around $4000$ MeV.

\begin{figure}
\begin{center}
\includegraphics{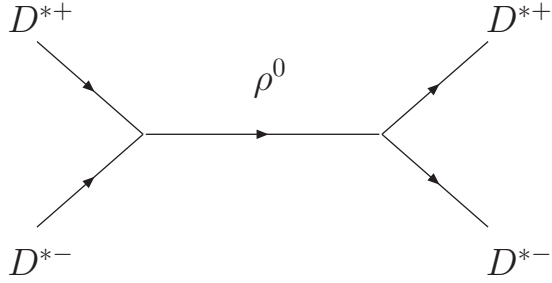}
\end{center}
\caption{$S$-channel $\rho$ exchange diagram.}
\label{fig:fig8}
\end{figure}

The formalism that we are using is also allowed for $s$-channel vector exchange 
and we can have the diagram of Fig.~\ref{fig:fig8}. But we found in
\cite{raquel}
that this leads to a $p$-wave interaction for equal masses of the
vectors, and only to a minor component of $s$-wave in the case of
different masses \cite{geng}, hence, we do not consider these terms. 

\section{T-matrix}
The results of the amplitudes discussed in the Section 2.2 and 2.3 
provide the kernel or potential $V$ to be used in the Bethe-Salpeter
equation 
in its on-shell factorized form,
\begin{equation}
T= (\hat{1}-VG)^{-1} V\ .
\label{Bethe}
\end{equation}
 The potencial $V$ here is a $10\times 10$ matrix in $I=0$ with the amplitudes obtained from the channels 
 \begin{center}
 $\mathbf{D^*\bar{D}^*}(4017)$, $\mathbf{D^*_s\bar{D}^*_s}(4225)$, $\mathbf{K^*\bar{K}^*}(1783)$, $\mathbf{\rho\rho}(1551)$, $\mathbf{\omega\omega}(1565)$\\\vspace{0.5cm}  $\mathbf{\phi\phi}(2039)$, $\mathbf{J/\psi J/\psi} (6194)$, $\mathbf{\omega J/\psi} (3880)$, $\mathbf{\phi J/\psi} (4116)$, $\mathbf{\omega \phi} (1802)$
 \end{center}
 in its elements for each spin $J=0,1,2$ independently. Also, in $I=1$, $V$ is a $6\times6$ matrix whose elements come from the transition potentials between the channels
 \begin{center}
$\mathbf{D^*\bar{D}^*}(4017)$, $\mathbf{K^*\bar{K}^*}(1783)$, $\mathbf{\rho\rho}(1551)$, $\mathbf{\rho\omega}(1558)$, $\mathbf{\rho J\psi}(3872)$, $\mathbf{\rho\phi} (1795)$.
\end{center}
In Eq. (\ref{Bethe}) $G$ is a diagonal matrix where its elements are the two meson loop 
function $G_i$ for each channel $i$:
\begin{equation}
G_i=i\int \frac{d^4 q}{(2\pi)^4}\frac{1}{q^2-m_1^2+i\eps}\frac{1}{(P-q)^2-m_2^2+i\eps}\ ,
\label{loop}
\end{equation}
which upon using dimensional regularization can be recast as
\begin{eqnarray}
G_i&=&{1 \over 16\pi ^2}\biggr( \alpha +Log{m_1^2 \over \mu ^2}+{m_2^2-m_1^2+s\over 2s}
  Log{m_2^2 \over m_1^2}\nonumber\\ 
  &+&{p\over \sqrt{s}}\Big( Log{s-m_2^2+m_1^2+2p\sqrt{s} \over -s+m_2^2-m_1^2+
  2p\sqrt{s}}+Log{s+m_2^2-m_1^2+2p\sqrt{s} \over -s-m_2^2+m_1^2+  2p\sqrt{s}}\Big)\biggr)\ ,
  \label{dimreg}
\end{eqnarray}
 where $P$ is the total four-momentum of the two mesons and $p$ is the three-momentum 
 of the mesons in the center-of-mass frame:
 \begin{equation}
 p=\frac{\sqrt{(s-(m_1+m_2)^2)\,(s-(m_1-m_2)^2)}}{2\,\sqrt{s}}\ .
 \end{equation}
 Analogously, using a cut 
 off one obtains
 \begin{equation}
G_i=\int_0^{q_{max}} \frac{q^2 dq}{(2\pi)^2} \frac{\omega_1+\omega_2}{\omega_1\omega_2 [{(P^0)}^2-(\omega_1+\omega_2)^2+i\epsilon]   } \ ,\label{loopcut}
\end{equation}
 where $q_{max}$ stands for the cut off in the three-momentum, $\omega_i=(\vec{q}\,^2_i+m_i^2)^{1/2}$ 
 and the square of center-of-mass energy ${(P^0)}^2=s$. In the complex plane, for a general $\sqrt{s}$, the loop function in the second Riemann sheet can be written as \cite{Roca}:
 \begin{equation}
G^{II}_i(\sqrt{s})=G^{I}_i(\sqrt{s})+i \frac{p}{4\pi \sqrt{s}}\,\hspace{1cm}Im(p)>0
\label{secondR}
\end{equation}
where $G^{II}_i$ refer to the loop function in the second Riemann sheet and $G^{I}_i$ is the loop function in the first Riemann sheet given by Eqs. (\ref{dimreg}) and (\ref{loopcut}) for each channel $i$.

\section{Results}

Once we introduce the potentials obtained from the Tables
\ref{tab:Am1}, \ref{tab:Am2}, \ref{tab:Am3}, \ref{tab:Am4}, \ref{tab:Am5} and
\ref{tab:Am6} (in the Appendix) as a kernel $V$ in Eq.~(\ref{Bethe}), we evaluate the transition matrix $T$ between channels and look at $|T|^2$ in the real axes and look for poles in the second Riemann sheet of the complex plane. If these poles are not very far from the real axis they occur in $\sqrt{s_{p}}=(M\pm i \Gamma/2)$. Since the only meaningful physical quantity is the value of the amplitudes for real $\sqrt{s}$, only poles not too far from the real axes would be easily identified experimentally as a resonance. The amplitude in Eq.~(\ref{Bethe}) close to a pole looks like
\begin{equation}
T_{ij}\approx \frac{g_i g_j}{s-s_{p}}\ ,
\label{poleT}
\end{equation}
 The constants $g_i$ ($i=channel$), which provide
the coupling of the resonance 
to one particular channel are calculated by means of the residues of
the amplitudes. 

We have set the parameters of Eq.~(\ref{dimreg}) as follow: we have fixed the value of
$\mu$ as $1000$ MeV for all the channels, the subtraction constant for the loops containing $SU(3)$ mesons,  that we call as $\alpha_L$, is set to $-1.65$ in order to find the position of the pole $f_2(1275)$ as in \cite{geng}, for the channels where the two mesons are heavy we put $\alpha_H=-2.07$, value that is chosen to get the position of the pole found in $S=0$ around $3940$ MeV, and for the rest of the channels, $\omega J/\psi$, $\phi J/\psi $, $\rho J\psi$, we also put $\alpha_L=-1.65$. One could do a study of the theoretical uncertainties for the positions and widths to account for underlying SU(4) symmetry breaking of the model. This study was done in \cite{raquel2} and one finds similar results here. In summary, using the freedom of the model to use $f_\pi$ or $f_D$ and $M_V$, but readjusting the substraction constants to obtain one of the resonances with the same mass as the experiment, one finds changes in the masses smaller than $20$ MeV and changes in the couplings smaller than $8\%$.

We find four poles in $I=0$ and one pole in $I=1$. The pole positions and the coupling constans $g_i$ are given in Tables
\ref{tab:res1}, \ref{tab:res2}, \ref{tab:res3}, \ref{tab:res4} and \ref{tab:res5}. In $I=0$ we have found one pole with a mass $\sim 3940$ MeV for each spin $J=0,1,2$ and from the couplings we see that they couple stronger to $D^*\bar{D}^*$. For $J=0$ and $J=1$, the width found is very small, $\Gamma_{p}=14.8$ and $0$ MeV respectively, but for $J=2$ the width is comparatively larger $\Gamma_{p}=52$ MeV. Another pole is also found for $I=0$ and $J=2$ with mass $M_{p}=4169$ MeV and width $\Gamma_p=132$ MeV which couples now stronger to $D^*_s\bar{D}^*_s$. In the $I=1$ sector, we find a pole only for $J=2$ with mass $M_p=3919$ and width $\Gamma_p=148$ that couples mostly to $D^*\bar{D}^*$. In \cite{Godfrey}, the XYZ mesons observed up to now are listed in Table 1. To make the appropiate correspondence, let us quote from \cite{Godfrey} the properties of the XYZ states discussed, which we show in Table \ref{tab:new}. 

\begin{table}[h]
\begin{center}
\begin{tabular}{c|c|c|c|c|c}
State & M (MeV) & $\Gamma$ (MeV) & $J^{PC}$ & Decay modes & Production modes \\
\hline
\hline
$Z(3940)$& $3929\pm5$ & $29\pm 10$ \T\B& $2^{++}$ & $D\bar{D}$ & $\gamma\gamma$ \\
\hline
$X(3940)$ & $3942\pm 9$ &$37\pm 17$ \T\B& $J^{P+}$ & $D\bar{D}^*$ & $e^+ e^-\to J/\psi X(3940)$\\
\hline
$Y(3940)$ & $3943\pm 17$  & $87\pm 34$\T\B& $J^{P+}$ & $\omega J/\psi$ & $B\to KY(3940)$\\
 & $3914.3^{+4.1}_{-3.8}$& $33^{+12}_{-8}$\T\B& & & \\
\hline
$X(4160)$ & $4156\pm 29$ & $139^{+113}_{-65}$\T\B& $J^{P+}$ & $D^{*}\bar{D}^*$&$e^+e^-\to J/\psi X(4160)$\\
\hline
\end{tabular}
\end{center}
\caption{Properties of the candidate XYZ mesons.}
\label{tab:new}
\end{table}

Looking at this table, we see that there are three states with mass around $3940$ MeV. The identification is subtle. At first one could think that the three states around $3940$ MeV could correspond to our states found around this mass. However, all the three experimental states have C-parity positive, while our state at $3945$ MeV has C-parity negative. Hence, this state is a prediction and does not correspond to any of the experimental ones. We call it $Y_p(3945)$ where $p$ stands for prediction. The other two states that appear in Table \ref{tab:new} with masses around $3940$ MeV are the $Z(3940)$ and the $Y(3940)$. The $Z(3940)$ has been seen in the spectrum of $D\bar{D}$ produced in $\gamma\gamma$ collisions with mass $M=(3929\pm 5)$ MeV and $\Gamma=(29\pm 10)$ MeV and the two-photon production process can only produce $D\bar{D}$ in a $0^{++}$ or $2^{++}$ state. From the angular distributions the Belle measurement strongly favors the $2^{++}$ hypothesis, and for this reason we associate this state to our $2^{++}$ state, which is found with mass and width $M_p=3922$ MeV and $\Gamma_p=52$ MeV. For the $Y(3940)$ there are different measurements of the mass and width, whereas Babar reports $M=(3914.3^{+4.1}_{-3.8})$ MeV and $\Gamma=(33^{+12}_{-8})$ MeV, the Belle's values are $M=(3943\pm 17)$ MeV and $\Gamma=(87\pm 34)$ MeV. Even though the difference in the values of the mass is not large, there is a considerable uncertainty in the width. This is the state that we associate to our $0^{++}$ state, which we find with mass $M_p=3943$ MeV and width $\Gamma_p=14.8$ MeV. The point in favor of this assignment is the result of the calculation of $\Gamma((3943,0^{+}[0^{++}])\to \omega J/\psi)$, that can be done straightforward by means the formula:
\begin{equation}
\Gamma((3943,0^{+}[0^{++}])\to \omega J/\psi)=\frac{p\,|g_{Y\omega J/\psi}|^2}{8 \pi\,M_Y^2} 
\label{decay}
\end{equation}
with $p$ the momentum of $\omega$ in the resonance rest frame. Taking the coupling $g_{Y\omega J/\psi}=(-1429-i\,216)$ MeV from Table \ref{tab:res1}, we obtain $\Gamma((3943,0^{+}[0^{++}])\to \omega J/\psi)=1.52$ MeV, which is compatible with $\Gamma(Y(3940)\to \omega J/\psi)>1$ MeV, obtained from the measured product of branching fractions $\mathcal{B}(B\to K Y(3940)) \mathcal{B}(Y(3940)\to \omega J/\psi)$ ($(7.1\pm 3.4)\times 10^{-5}$, reported by Belle, and  $(4.9\pm 1.1)\times 10^{-5}$ according to Babar) together with the assumption that $\mathcal{B}(B\to KY(3940))$ is less than or equal to $1\times 10^{-3}$, the typical value for allowed $B\to K+$ charmonium decays. Thus, we find a natural explanation on why this rate is much larger than it would be should it correspond to hadronic transitions between charmonium states \cite{Godfrey}.

The state $X(3940)$ of Table \ref{tab:new} does not decay into $\omega J/\psi$ \cite{Abe2}. This is a reason not to associate it to our $(3943, 0^+[0^{++}])$. This state does not fit into our vector - vector scheme, suggesting it has a different nature.

The state found with $J^{PC}=2^{++}$, $M_p=4169$ MeV and $\Gamma_p=132$ MeV can be clearly identified with the $X(4160)$ by the proximity of mass, width and quantum numbers.

In the next section, we are going to include the $D\bar{D}$ channel by means a box diagram as it was done in \cite{raquel,raquel2,geng}.

\begin{table}[h]
 \begin{center}
\centerline{$\sqrt{s}_{pole}=3943 + i 7.4$, $I^G[J^{PC}]=0^+[0^{++}]$}
\vspace{0.5cm}
\begin{tabular}{ccccc}
\hline
$D^*\bar{D}^*$&$D^*_s\bar{D}_s^*$&$K^*\bar{K}^*$&$\rho\rho$&$\omega\omega$\\
\hline
\hline
$18810-i 682 $&$8426+i 1933 $&$10- i 11$&$-22 + i 47$&$1348+ i 234 $\\
\hline
\end{tabular}\\
\vspace{0.5cm}
\begin{tabular}{ccccc}
\hline
$\phi\phi$&$J/\psi J/\psi$&$\omega J/\psi$&$\phi J/\psi$&$\omega\phi$\\
\hline
\hline
$-1000 -i 150$&$417+ i 64$&$-1429 - i 216$&$889+ i 196 $&$-215 - i107$\\
\hline
\end{tabular} 
\end{center}
\caption{Couplings $g_{i}$ in units of MeV for $I=0$, $J=0$.}
\label{tab:res1}
\end{table} 
\begin{table}[h]
\begin{center}
\centerline{$\sqrt{s}_{pole}=3945 +i 0$, $I^G[J^{PC}]=0^-[1^{+-}]$}
\vspace{0.4cm}
\begin{tabular}{cccccccccc}
\hline
$D^*\bar{D}^*$&$D^*_s\bar{D}_s^*$&$K^*\bar{K}^*$&$\rho\rho$&$\omega\omega$&$\phi\phi$&$J/\psi J/\psi$&$\omega J/\psi$&$\phi J/\psi$&$\omega\phi$\\
\hline
\hline
$18489- i0.78 $&$8763+ i2 $&$11-i38 $&$0$&$0$&$0$&$0$&$0$&$0$&$0$\\
\hline
\end{tabular}
\end{center}
\caption{Couplings $g_{i}$ in units of MeV for $I=0$, $J=1$.}
\label{tab:res2}
\end{table} 
\begin{table}[h]
\begin{center}
\centerline{$\sqrt{s}_{pole}=3922+i 26$, $I^G[J^{PC}]=0^+[2^{++}]$}
\vspace{0.5cm}
\begin{tabular}{ccccc}
\hline
$D^*\bar{D}^*$&$D^*_s\bar{D}_s^*$&$K^*\bar{K}^*$&$\rho\rho$&$\omega\omega$\\
\hline
\hline
$21100- i 1802 $&$1633+i 6797 $&$42+ i 14 $&$-75 +i 37$&$1558 + i 1821$\\
\hline
\end{tabular}\\
\vspace{0.5cm}
\begin{tabular}{ccccc}
\hline
$\phi\phi$&$J/\psi J/\psi$&$\omega J/\psi$&$\phi J/\psi$&$\omega\phi$\\
\hline
\hline
$-904 - i1783 $&$1783 +i 197$&$-2558 - i2289$&$918+ i2921 $&$91 -i 784$\\
\hline
\end{tabular}
\end{center}
\caption{Couplings $g_{i}$ in units of MeV for $I=0$, $J=2$.}
\label{tab:res3}
\end{table} 
\begin{table}[h]
\begin{center}
\centerline{$\sqrt{s}_{pole}=4169+i 66$, $I^G[J^{PC}]=0^+[2^{++}]$}
\vspace{0.5cm}
\begin{tabular}{ccccc}
\hline
$D^*\bar{D}^*$&$D^*_s\bar{D}_s^*$&$K^*\bar{K}^*$&$\rho\rho$&$\omega\omega$\\
\hline
\hline
$1225- i 490 $&$18927- i 5524$&$-82 + i 30 $&$70+ i20$&$3 -i 2441$\\
\hline
\end{tabular}\\
\vspace{0.5cm}
\begin{tabular}{ccccc}
\hline
$\phi\phi$&$J/\psi J/\psi$&$\omega J/\psi$&$\phi J/\psi$&$\omega\phi$\\
\hline
\hline
$1257+ i 2866 $&$2681+ i 940$&$-866 + i 2752 $&$-2617 - i5151 $&$1012+ i 1522$\\
\hline
\end{tabular}
\end{center}
\caption{Couplings $g_{i}$ in units of MeV for $I=0$, $J=2$ (second pole).}
\label{tab:res4}
\end{table} 
\begin{table}[h]
\begin{center}
\centerline{$\sqrt{s}_{pole}=3919+ i74$, $I^G[J^{PC}]=1^-[2^{++}]$}
\vspace{0.4cm}
\begin{tabular}{cccccc}
\hline
$D^*\bar{D}^*$&$K^*\bar{K}^*$&$\rho\rho$&$\rho\omega$&$\rho J/\psi$&$\rho\phi$\\
\hline
\hline
$20267-i 4975 $&$148- i33 $&$0$&$-1150 -i 3470 $&$2105+ i5978$&$-1067 -i 2514$\\
\hline
\end{tabular}
\end{center}
\caption{Couplings $g_{i}$ in units of MeV for $I=1$, $J=2$.}
\label{tab:res5}
\end{table}

\section{The $D \bar{D}$ decay mode}
\subsection{Evaluation of the $D\bar{D}$-box diagram}

\begin{figure}
\centering
\includegraphics[width=13cm]{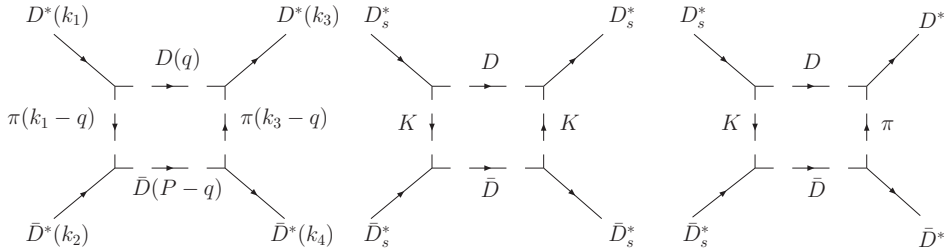} \\
\caption{$D\bar{D}$-box diagrams for the $D^*\bar{D}^*$ and $D^*_s\bar{D}^*_s$ channels.}
\label{fig:fig9}
\end{figure}

Here we consider the diagrams of Fig.~\ref{fig:fig9} in order to take into account the $D\bar{D}$ decay that we introduce in the most important channels: $D^*\bar{D}^*$ and $D^*_s\bar{D}^*_s$. Our starting point is the set of
diagrams of 
Fig.~\ref{fig:fig9}. Take the first one, suppose $D^*=D^{*+}$, $\bar{D}^*=D^{*-}$ and $\pi=\pi^0$, the vextex are provided within the same hidden gauge formalism
\cite{hidden1}, \cite{hidden2}, used in Section $2$, by means of the
Lagrangian 
\begin{equation}
{\cal L}_{V\Phi\Phi}=-ig\langle V^\mu[\Phi,\partial_\mu \Phi]\rangle\ .
\label{lVPP}
\end{equation}
For these particles, we have, for the first diagram of Fig.~\ref{fig:fig9}:
\begin{eqnarray}
-it^{(D\bar{D})}&=&\int\frac{d^4 q}{(2\pi)^4}(-i)^4\,g_D^4\,(\frac{1}{\sqrt{2}})^4\,  (k_1-2q)^\mu\eps^{(1)}_\mu \nonumber\\
&\times& (k_3-2q)^\nu\eps^{(3)}_{\nu} (P+k_1-2q)^\alpha\eps^{(2)}_{\alpha}
(P+k_3-2q)^\beta\eps^{(4)}_{\beta}\nonumber\\&\times&\frac{i}{q^2-m^2_D+i\eps}\,\frac{i}{(k_1-q)^2-m^2_\pi+i\eps}\nonumber\\
&\times&\,\frac{i}{(P-q)^2-m^2_D+i\eps}\,\frac{i}{(k_3-q)^2-m^2_\pi+i\eps}\ .
\label{tbox}
\end{eqnarray}
Using the approximation that all the polarization vectors are spatial,
it is possible to write the above amplitude as 
\begin{eqnarray}
-it^{(D\bar{D})}&=&g_D^4\int\frac{d^4q}{(2\pi)^4}4\,q_i q_jq_lq_m \, \eps^{(1)}_i\eps^{(2)}_j\eps^{(3)}_l\eps^{(4)}_m\nonumber\\&\times&\frac{1}{q^2-m^2_D+i\eps}\,\frac{1}{(k_1-q)^2-m^2_\pi+i\eps}\nonumber\\
&\times&\frac{1}{(P-q)^2-m^2_D+i\eps}\,\frac{1}{(k_3-q)^2-m^2_\pi+i\eps}\ .
\label{tbox1}
\end{eqnarray}
This integral is logarithmically divergent and as in \cite{raquel}
we regularize it with a cutoff in the three-momentum of the order of the
natural size,
for which we take
$q_{max}=1.2\,GeV$.
The results do not
change much if one takes a value around this. 
After performing analytically the $dq^0$ integral of Eq.~(\ref{tbox1}),
one finds 
\begin{eqnarray}
V^{(D \bar{D})}&=&g_D^4\, \left( \eps^{(1)}_i\eps^{(2)}_i\eps^{(3)}_j\eps^{(4)}_j+\eps^{(1)}_i\eps^{(2)}_j\eps^{(3)}_i\eps^{(4)}_j+\eps^{(1)}_i\eps^{(2)}_j\eps^{(3)}_j\eps^{(4)}_i\right)\nonumber\\
&\times& \frac{2}{15\pi^2}\int^{q_{max}}_0\,dq\, \vec{q}\,^6\,\left(\frac{1}{\omega}\right)^3\frac{1}{\omega_D}\left( \frac{1}{k_3^0+\omega+\omega_D}\right) ^2\frac{1}{k_1^0-\omega-\omega_D+i\eps}\nonumber\\
&\times&\frac{1}{k_3^0-\omega-\omega_D+i\eps}\,
\frac{1}{P^0-2\omega_D+i\eps}\,\frac{1}{P^0+2\omega_D}\nonumber\\
&\times& \left(\omega^3+\omega_D^3+4\,\omega\omega_D^2+4\omega^2\omega_D-k^0_3\omega_D\right)
\label{Vbox}
\end{eqnarray}
where $\omega=\sqrt{\vec{q}\,^2+ m_\pi ^2}$, $\omega_D=\sqrt{\vec{q}\,^2+
m_D ^2}$, $P^0=k^0_1+k^0_2$. In Eq.~(\ref{Vbox}) we can see clearly the
sources of the imaginary part in the cuts
$k_1^0(k_3^0)-\omega-\omega_D=0$, $P^0-2\omega_D=0$, which give rise to the decays $D^{*+}\to \pi^0
D^+$ and $D^{*+} D^{*-}\to D^+ D^-$ respectively. 
%
After isospin and spin projection, we obain 
\begin{eqnarray}
t^{(D\bar{D},I=0,J=0)}&=&45\,\tilde{V}^{(D\bar{D})}\nonumber\\
t^{(D\bar{D},I=0,J=2)}&=&18\,\tilde{V}^{(D\bar{D})}
\label{eq:boxsis}
\end{eqnarray}
and
\begin{eqnarray}
t^{(D\bar{D},I=1,J=0)}&=&5\,\tilde{V}^{(D\bar{D})}\nonumber\\
t^{(D\bar{D},I=1,J=2)}&=&2\,\tilde{V}^{(D\bar{D})}\ ,
\label{eq:boxsis1}
\end{eqnarray}
where $\tilde{V}^{(D\bar{D})}$ is given by Eq.~(\ref{Vbox}) after removing the
polarization vectors. That is, we have found that the $D\bar{D}$ decay channel contributes only to $J=0$ and $J=2$. The case $J=1$ is forbidden, as already found in \cite{raquel,raquel2}. This is 
because
the parity of the $D^*\bar{D}^*$
system for s-wave is positive, which forces the $D\bar{D}$
system to be in $L=0,2$. Since the $D$ and $\bar{D}$ have no spin, the total angular momentum $J$ is equal to $L$ in this case. Therefore, only the $0^+$, $2^+$ quantum
numbers have this decay channel.

\begin{figure}
\begin{center}
\includegraphics[width=16.5cm, angle=0]{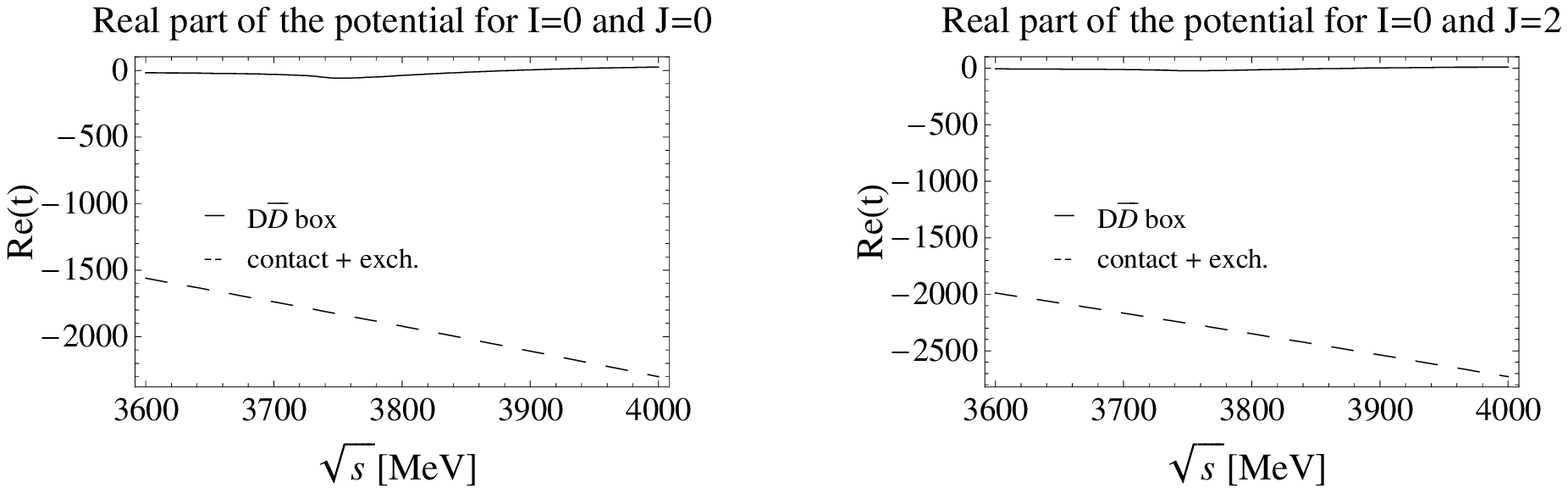} \\ 
\end{center}
\caption{Real parts of the potential for the $D^*\bar{D}^*$ channel.}
\label{fig:fig10}
\end{figure}

\begin{figure}
\begin{center}
\includegraphics[width=16.5cm, angle=0]{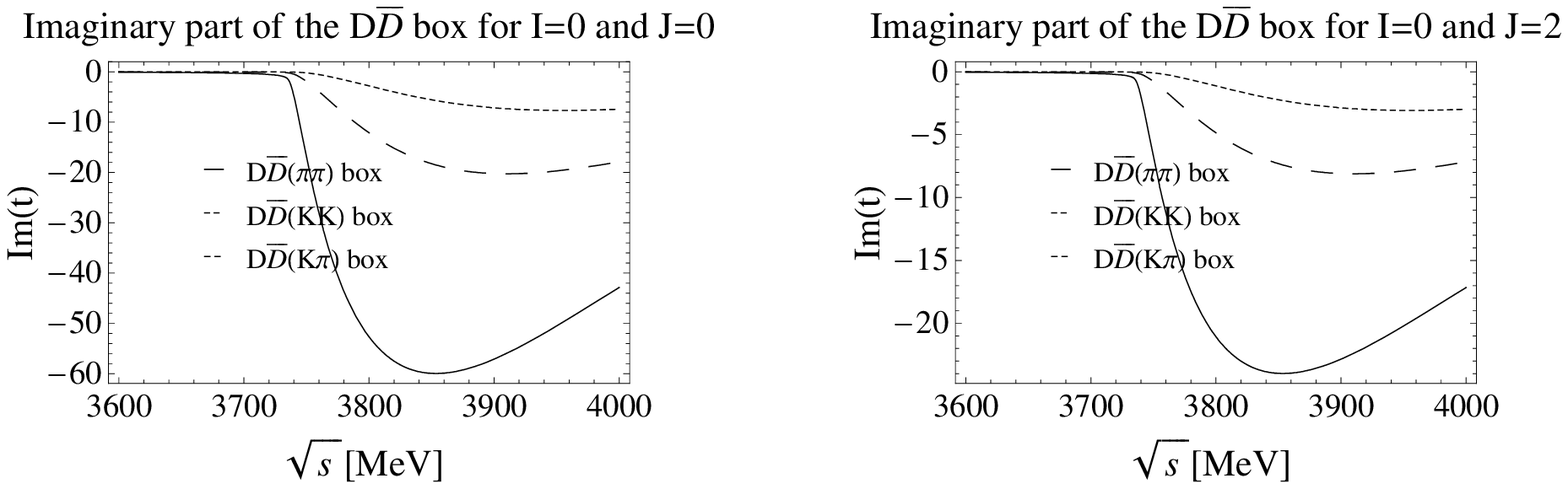} \\ 
\end{center}
\caption{Imaginary parts of the $D\bar{D}$ box diagrams of Fig. \ref{fig:fig9}.}
\label{fig:fig11}
\end{figure}

For the second diagram of Fig. \ref{fig:fig9}, the evaluation of the integral is very similar to that of Eq. (\ref{tbox1}), and it becomes the formula of Eq. (\ref{Vbox}) except for a factor and changing $m_\pi\to m_K$ and $\omega\to \omega_K$, being $\omega_K=\sqrt{\vec{q}\,^2+m^2_K}$, that is

\begin{eqnarray}
t_K^{(D\bar{D},I=0,J=0)}&=&10\,\tilde{V}_K^{(D\bar{D})}\nonumber\\
t_K^{(D\bar{D},I=0,J=2)}&=&4\,\tilde{V}_K^{(D\bar{D})}
\label{eq:boxsisK}
\end{eqnarray}
with 
\begin{eqnarray}
\tilde{V}_K^{(D \bar{D})}&=&g_{D_s}^4\, 
\frac{8}{15\pi^2}\int^{q_{max}}_0\,dq\, \vec{q}\,^6\,\left(\frac{1}{\omega_K}\right)^3\frac{1}{\omega_D}\left( \frac{1}{k_3^0+\omega_K+\omega_D}\right) ^2\nonumber\\
&\times&\frac{1}{k_1^0
-\omega_K-\omega_D+i\eps}\,\frac{1}{k_3^0
-\omega_K-\omega_D+i\eps}\,
\frac{1}{P^0-2\omega_D+i\eps}\nonumber\\
&\times& \frac{1}{P^0+2\omega_D}\,\left(\omega_K^3+\omega_D^3+4\,\omega_K\omega_D^2+4\omega_K^2\omega_D-k^0_3\omega_D\right)
\label{VboxK}
\end{eqnarray}
For the third diagram of Fig. \ref{fig:fig9}, we obtain
\begin{eqnarray}
t_{K\pi}^{(D\bar{D},I=0,J=0)}&=&\frac{15}{\sqrt{2}}\,\tilde{V}_{K\pi}^{(D\bar{D})}\nonumber\\
t_{K\pi}^{(D\bar{D},I=0,J=2)}&=&\frac{6}{\sqrt{2}}\,\tilde{V}_{K\pi}^{(D\bar{D})}
\label{eq:boxsisKpi}
\end{eqnarray}
with
\begin{eqnarray}
\tilde{V}_{K\pi}^{(D \bar{D})}&=&g_{D_s}^2 g_{D}^2\, 
\frac{8}{15\pi^2}\int^{q_{max}}_0\,dq\, \vec{q}\,^6\,\frac{1}{\omega\,\omega_K\,\omega_D}\,\frac{1}{\omega+\omega_K} \frac{1}{k_3^0+\omega+\omega_D} \,\frac{1}{k_1^0+\omega_K+\omega_D}\nonumber\\
&\times&\frac{1}{k_1-\omega_K-\omega_D+i\eps}\,\frac{1}{k_3^0-\omega-\omega_D+i\eps}\,
\frac{1}{P^0-2\omega_D+i\eps}\,\frac{1}{P^0+2\omega_D}\nonumber\\
&\times& \left( 2\omega_D(\omega_D+\omega_K)^2+\omega^2(2\omega_D+\omega_K)+\omega(2\omega_D+\omega_K)^2-2 k^{0\,2}_3\omega_D\right)\ .
\label{VboxKpi}
\end{eqnarray}
 As in \cite{raquel2} we use a form factor for an
off-shell $\pi(K)$ in each vertex, which is 
\begin{equation}
F(q)=e^{-\frac{\vec{q}\,^2}{\Lambda^2}}
\label{formfactor}
\end{equation}
with $\Lambda=1.2$ GeV \cite{Navarra}. The
real part of the potential coming from the $D\bar{D}$ box is much
smaller than the real part of the potential coming from the contact plus
exchange terms as we can see in the representative figure ,Fig. \ref{fig:fig10}, for the $D^*\bar{D}^*$ channel. Then, it can be neglected and we keep only the imaginary parts ot the $D\bar{D}$ box diagrams, which are plotted in Fig. \ref{fig:fig11}. In this figure, we see that the most important contribution comes from the $D\bar{D}(\pi\pi)$ box diagram, the $D\bar{D}(KK)$ and $D\bar{D}(K\pi)$ contributions being less relevant.

\subsection{Results with $V^{(D\bar{D})}$}

\begin{figure}
\begin{center}
\begin{tabular}{cc}
\includegraphics[width=7cm, angle=0]{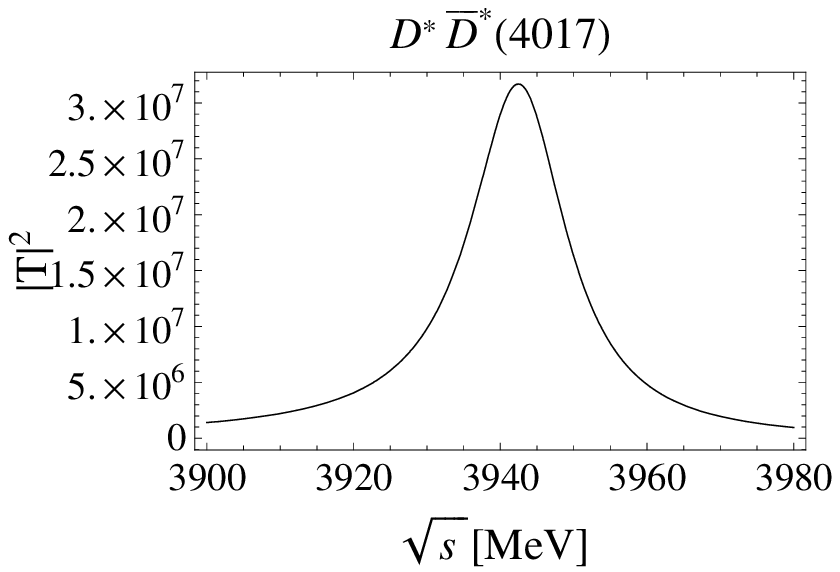} &
\includegraphics[width=7cm, angle=0]{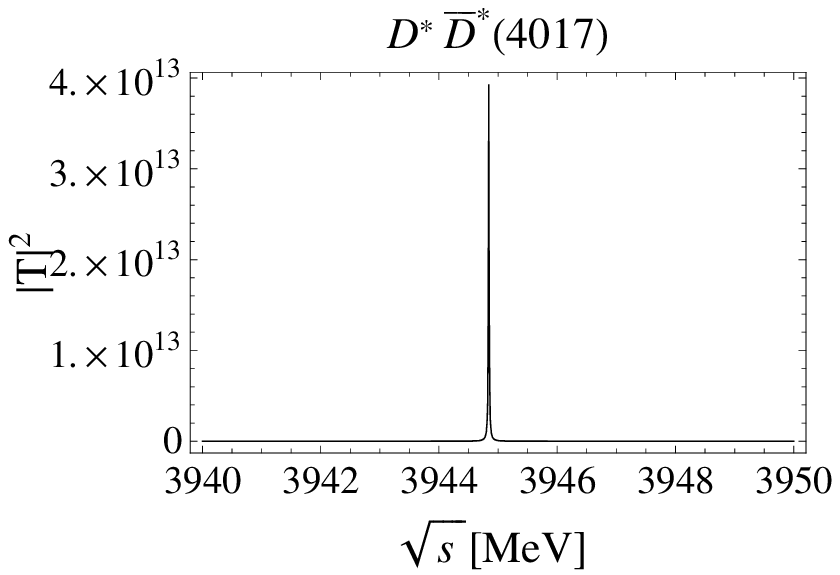}
\end{tabular}
\end{center}
\caption{$|T|^2$ for $I=0$ and $J=0$ (left), $J=1$ (right), in the main channel $D^*\bar{D}^*$.}
\label{T001}
\end{figure}

\begin{figure}
\begin{center}
\begin{tabular}{cc}
\includegraphics[width=7cm, angle=0]{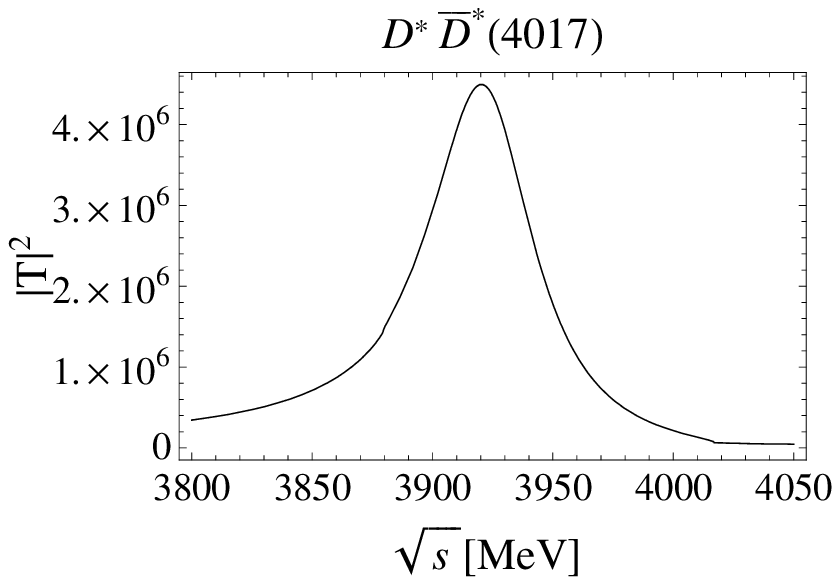} &
\includegraphics[width=7cm, angle=0]{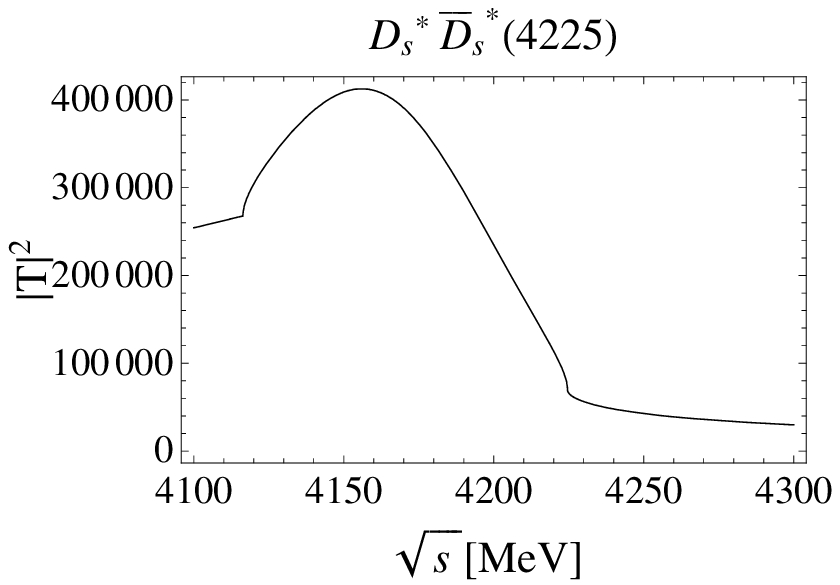}
\end{tabular}
\end{center}
\caption{$|T|^2$ for $I=0$ and $J=2$ in the main channels $D^*\bar{D}^*$ (first pole, left) and $D^*_s \bar{D}^*_s$ (second pole, right).}
\label{T02}
\end{figure}

\begin{figure}
\begin{center}
\includegraphics[width=7cm, angle=0]{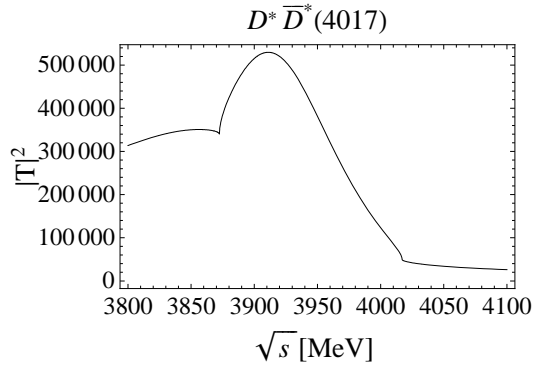} \\ 
\end{center}
\caption{$|T|^2$ for $I=1$ and $J=2$ in the main channel $D^*\bar{D}^*$.}
\label{T12}
\end{figure}

\begin{figure}
\begin{center}
\includegraphics[width=9cm, angle=0]{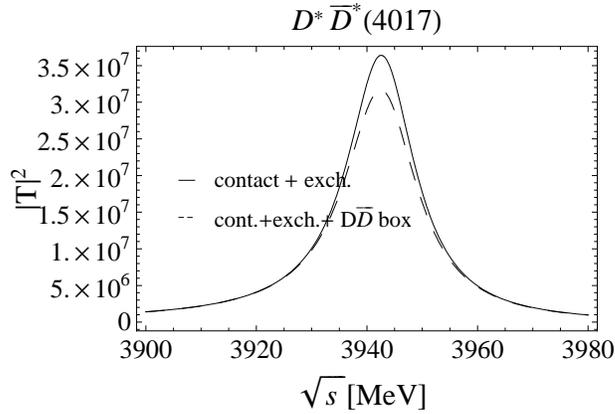} \\ 
\end{center}
\caption{$|T|^2$ for $I=0$ and $J=0$ before and after the inclusion of the $D\bar{D}$ box diagrams.}
\label{T00b}
\end{figure}

In Figs.~\ref{T001}, \ref{T02} and \ref{T12} we show the results when one introduces the
amplitudes obtained in Section 2 and the $D\bar{D}$ box diagrams of Fig. \ref{fig:fig9} in the Bethe-Salpeter
equation Eq. (\ref{Bethe}). The masses and widths in the real axis obtained from these plots are given in Table \ref{tab:exp}. As one can see, the results for the mass and widths of the resonances have not changed too much with respect to those of Section 4, the inclusion of the $D\bar{D}$ diagrams modifies only a few MeV the final width. The effect can be seen in Fig. \ref{T00b} for the resonance in the $I=0$, $J=0$ sector. Thus, the main decay channels of our states are the light vector - light vector decay modes as $K^*\bar{K}^*$, $\rho\rho$, $\omega\omega$, $\phi\phi$, $\omega\phi$ in $I=0$, and $\rho\omega$, $\rho\phi$ for $I=1$, and the light vector - heavy vector decay modes, $\omega J/\psi$ for $I=0$, and $\rho J/\psi$ in $I=1$.

\begin{table}[htb!]
\begin{center}
\begin{tabular}{c|c|c|c|c|c|c}
$I^G[J^{PC}]$&\multicolumn{2}{c|}{Theory}&\multicolumn{4}{c}{Experiment}\\
\hline\hline
& Mass [MeV] & Width [MeV]\T\B & Name & Mass [MeV] & Width [MeV] &$J^{PC}$\\
$0^+[0^{++}]$&$3943$&$17$&$Y(3940)$&$3943\pm 17$&$87\pm 34$&$J^{P+}$\\
& & & & $3914.3^{+4.1}_{-3.8}$ & $33^{+12}_{-8}$ & \\
$0^-[1^{+-}]$&$3945$&$0$&"$Y_p(3945)$"& & & \\
$0^+[2^{++}]$&$3922$&$55$&$Z(3930)$&$3929\pm 5$& $29\pm 10$&$2^{++}$\\
$0^+[2^{++}]$&$4157$&$102$&$X(4160)$&$4156\pm 29$& $139^{+113}_{-65}$& $J^{P+}$\\
$1^-[2^{++}]$&$3912$&$120$&"$Y_p(3912)$"& & & \\
\hline
\end{tabular}
\end{center}
\caption{Comparison of the mass, width and quantum numbers with the experiment.}
\label{tab:exp}
\end{table}

\section{Conclusions}

We have made a full study of the vector - vector interaction in the $C=0$ and $S=0$ sector using the hidden gauge
formalism. The interaction comes from contact terms plus vector meson exchange
in the $t$-channel. We have found a strong attraction in the $I=0$, $J=0,1,2$ and $I=1$, $J=2$ sectors, enough to bind the vector - vector system. By looking for poles in the second Riemann sheet, we have found five resonances, three of which can be associated with the experimental data: The state $(3943,0^+[0^{++}])$ to the $Y(3940)$, the $(3922,0^+[2^{++}])$ to the $Z(3930)$ and the $(4157,0^+[2^{++}])$ to the $X(4160)$. There is no experimental counterpart for our state $(3945,0^-[1^{+-}])$, which is thus a prediction of our model.  These three states with mass around $3940$ MeV are basically composed by $D^*\bar{D}^*$, and  decay into pairs of light vectors like $K^*\bar{K}^*$, or light vector - heavy vector as $\omega J/\psi$.
Our model predicts another state around $4160$ MeV, $(4157,0^+[2^{++}])$, which we identify with the $X(4160)$ state in base to the proximity of mass and widths and C-parity. This resonance has $J^{PC}=2^{++}$ and is mostly $D^*_s\bar{D}^*_s$. In the $I=1$ sector, the attraction is weak and we find only one resonance in the case of $J=2$, $(3912,1^-[2^{++}])$, the possible association of this state to the $X(3940)$ is unlikely since our state can decay to $D\bar{D}$, though with small intensity, but this decay is not seen for the $X(3940)$. The width that we obtain is also considerably larger than that of the $X(3940)$, $\Gamma=120$ MeV. 

As we have shown in this work, the region around $3940$ MeV is very interesting and there could be more resonances not yet seen in this region. The findings of this work should motivate the experimentalist to look into this region in the channels that involve light vector - light vector or light vector - heavy vector like $K^*\bar{K}^*$ and $\rho J/\psi$.

\section*{Acknowledgments}

This work is partly supported by DGICYT contract number
FIS2006-03438. We acknowledge the support of the European Community-Research Infrastructure
Integrating Activity
Study of Strongly Interacting Matter (acronym HadronPhysics2, Grant Agreement
n. 227431)
under the Seventh Framework Programme of EU.

\section*{Appendix}
\begin{table}[h]
\begin{tabular}{c|c|c|c}
 &Contact & V-exchange & $\sim Total[I[J^P]]$\\
\hline\hline
 \T\B$D^*\bar{D}^*\to D^*\bar{D}^*$& $6\,g_D^2$&$\frac{g_D^2\,(2 M_\omega^2 M_\rho^2+M_{J/\psi}^2 (3 M_\omega^2+M_\rho^2)) (4 M_{D^*}^2-3 s)}{4\, M_{J/\psi}^2 M_\omega^2 M_\rho^2} $& $-49.1\,g_D^2$ \\\T\B
$D^*\bar{D}^*\to D_s^*\bar{D}_s^*$& $2\sqrt{2}\,g_D g_{D_s}$& $\frac{g_D g_{D_s}(2M_{D^*}^2+2M_{D^*_s}^2-3 s)}{\sqrt{2}\,M^2_{K^*}}$& $-25.1\,g_Dg_{D_s}$\\\T\B
$D^*\bar{D}^*\to K^*\bar{K}^*$&$-2\,g g_D$ &$-\frac{g g_D (2M^2_{D^*}+2M^2_{K^*}-3s)}{2M^2_{D^*_s}}$ & $2.3\,g g_D$\\\T\B
$D^*\bar{D}^*\to \rho\rho$&$-2\sqrt{3}\,g g_D$ &$-\frac{\sqrt{3} g g_D (2 M^2_{D^*}+2 M^2_\rho-3s)}{2\,M^2_{D^*_s}}$ &$4.9\,g g_D$ \\\T\B
$D^*\bar{D}^*\to \omega\omega$& $2\, g g_D$& $\frac{g g_D (2M^2_{D^*}+2M^2_\omega-3s)}{2\,M^2_{D^*}}$& $-2.8\,g g_D$\\\T\B
$D^*\bar{D}^*\to \phi\phi$&$0$ & $0$& $0$\\\T\B
$D^*\bar{D}^*\to J/\psi J/\psi$&$4\,g_D g_{\eta_c}$ & $\frac{g_D g_{\eta_c} (2M^2_{D^*}+2M^2_{J/\psi}-3s)}{M^2_{D^*}}$& $-1.2\,g_D g_{\eta_c}$\\\T\B
$D^*\bar{D}^*\to \omega J/\psi$&$-4\,g g_D$ &$-\frac{g g_D (2M^2_{D^*}+M^2_{J/\psi}+M^2_\omega-3s)}{M^2_{D^*}}$ & $3.5\,g g_D$\\\T\B
$D^*\bar{D}^*\to \phi J/\psi$& $0$&$0$ &$0$ \\\T\B
$D^*\bar{D}^*\to \omega\phi$& $0$&$0$ &$0$ \\\T\B
$D_s^*\bar{D}_s^*\to D_s^*\bar{D}_s^*$& $4\,g^2_{D_s}$& $\frac{g^2_{D_s}(M^2_{J/\psi}+M^2_\phi) (4M^2_{D^*_s}-3s)}{2\,M^2_{J/\psi}M^2_\phi}$& $-12.3\,g^2_{D_s}$\\\T\B
$D_s^*\bar{D}_s^*\to K^*\bar{K}^*$& $-2\sqrt{2}\,g g_{D_s}$& $-\frac{g g_{D_s} (2M^2_{D^*_s}+2M^2_{K^*}-3s)}{\sqrt{2}M^2_{D^*}}$& $3.8\,g g_{D_s}$\\\T\B
$D_s^*\bar{D}_s^*\to \rho\rho$&$0$ &$0$ &$0$ \\\T\B
$D_s^*\bar{D}_s^*\to \omega\omega$&$0$ &$0$ &$0$ \\\T\B
$D_s^*\bar{D}_s^*\to \phi\phi$&$2\sqrt{2}\,g g_{D_s}$ & $\frac{g g_{D_s}(2M^2_{D^*_s}+2M^2_{\phi}-3s)}{\sqrt{2}M^2_{D^*_s}}$& $-3\,g g_{D_s}$\\\T\B
$D_s^*\bar{D}_s^*\to J/\psi J/\psi$& $2\sqrt{2}\,g_{D_s} g_{\eta_c}$& $\frac{g_{D_s} g_{\eta_c}(2M^2_{D^*_s}+2M^2_{J/\psi}-3s)}{\sqrt{2}M^2_{D^*_s}}$& $-0.4\,g_{D_s} g_{\eta_c}$\\
\T\B$D_s^*\bar{D}_s^*\to \omega J/\psi$& $0$& $0$&$0$ \\\T\B
$D_s^*\bar{D}_s^*\to \phi J/\psi$&$-4\,g g_{D_s}$ &$-\frac{g g_{D_s}(2M^2_{D^*_s}+M^2_{J/\psi}+M^2_\phi-3s)}{M^2_{D^*_s}}$ &$2.5\,g g_{D_s}$ \\\T\B
$D_s^*\bar{D}_s^*\to \omega\phi$&$0$ & $0$&$0$ \\
\end{tabular}
\caption{Amplitude projected in isospin and spin for $I=0$ and $J=0$. The approximate Total is obtained at the threshold of $D^*\bar{D}^*$.}
\label{tab:Am1}
\end{table}
\begin{table}[htb!]
\begin{tabular}{c|c|c|c}
 &Contact & V-exchange & $\sim Total[I[J^P]]$\\
\hline\hline
\T\B$D^*\bar{D}^*\to D^*\bar{D}^*$& $9\,g_D^2$ & $\frac{g_D^2\,(2 M_\omega^2 M_\rho^2+M_{J/\psi}^2 (3 M_\omega^2+M_\rho^2)) (4 M_{D^*}^2-3 s)}{4\, M_{J/\psi}^2 M_\omega^2 M_\rho^2} $ & $-46.1\,g_D^2$\\\T\B
$D^*\bar{D}^*\to D_s^*\bar{D}_s^*$& $3\sqrt{2}\,g_D g_{D_s}$& $\frac{g_D g_{D_s}(2M_{D^*}^2+2M_{D^*_s}^2-3 s)}{\sqrt{2}\,M^2_{K^*}}$& $-23.7\,g_D g_{D_s}$\\\T\B
$D^*\bar{D}^*\to K^*\bar{K}^*$&$3\,g g_D$ &$\frac{g g_D(2M^2_{D^*}+2M^2_{K^*}-3s)}{2M^2_{D^*_s}}$ & $-1.3\,g g_D$\\\T\B
$D^*\bar{D}^*\to \rho\rho$&$0$ &$0$ & $0$\\\T\B
$D^*\bar{D}^*\to \omega\omega$& $0$& $0$& $0$\\\T\B
$D^*\bar{D}^*\to \phi\phi$&$0$ & $0$& $0$\\\T\B
$D^*\bar{D}^*\to J/\psi J/\psi$&$0$ & $0$& $0$\\\T\B
$D^*\bar{D}^*\to \omega J/\psi$&$0$ &$0$ & $0$\\\T\B
$D^*\bar{D}^*\to \phi J/\psi$& $0$&$0$ &$0$ \\\T\B
$D^*\bar{D}^*\to \omega\phi$& $0$&$0$ &$0$ \\\T\B
$D_s^*\bar{D}_s^*\to D_s^*\bar{D}_s^*$& $6\,g^2_{D_s}$& $\frac{g^2_{D_s}(M^2_{J/\psi}+M^2_\phi) (4M^2_{D^*_s}-3s)}{2\,M^2_{J/\psi}M^2_\phi}$& $-10.3\,g_{D_s}^2$\\\T\B
$D_s^*\bar{D}_s^*\to K^*\bar{K}^*$& $-3\sqrt{2}\,g g_{D_s}$& $-\frac{g g_{D_s} (2M^2_{D^*_s}+2M^2_{K^*}-3s)}{\sqrt{2}M^2_{D^*}}$& $2.4\,g g_{D_s}$\\\T\B
$D_s^*\bar{D}_s^*\to \rho\rho$&$0$ &$0$ &$0$ \\\T\B
$D_s^*\bar{D}_s^*\to \omega\omega$&$0$ &$0$ &$0$ \\\T\B
$D_s^*\bar{D}_s^*\to \phi\phi$&$0$ & $0$& $0$\\\T\B
$D_s^*\bar{D}_s^*\to J/\psi J/\psi$& $0$& $0$&$0$ \\\T\B
$D_s^*\bar{D}_s^*\to \omega J/\psi$& $0$& $0$&$0$ \\\T\B
$D_s^*\bar{D}_s^*\to \phi J/\psi$&$0$ &$0$ & $0$\\\T\B
$D_s^*\bar{D}_s^*\to \omega\phi$&$0$ & $0$&$0$ \\
\end{tabular}
\caption{Amplitude projected in isospin and spin for $I=0$ and $J=1$.The approximate Total is obtained at the threshold of $D^*\bar{D}^*$. }
\label{tab:Am2}
\end{table}
\begin{table}[htb!]
\begin{tabular}{c|c|c|c}
 &Contact & V-exchange & $\sim Total[I[J^P]]$\\
\hline\hline
\T\B$D^*\bar{D}^*\to D^*\bar{D}^*$& $-3\,g_D^2$ & $\frac{g_D^2\,(2 M_\omega^2 M_\rho^2+M_{J/\psi}^2 (3 M_\omega^2+M_\rho^2)) (4 M_{D^*}^2-3 s)}{4\, M_{J/\psi}^2 M_\omega^2 M_\rho^2} $ & $-58.1\,g^2_D$\\\T\B
$D^*\bar{D}^*\to D_s^*\bar{D}_s^*$& $-\sqrt{2}\,g_D g_{D_s}$& $\frac{g_D g_{D_s}(2M_{D^*}^2+2M_{D^*_s}^2-3 s)}{\sqrt{2}\,M^2_{K^*}}$& $-29.4\,g_D g_{D_s}$\\\T\B
$D^*\bar{D}^*\to K^*\bar{K}^*$&$g g_D$ &$-\frac{g g_D(2M^2_{D^*}+2M^2_{K^*}-3s)}{2M^2_{D^*_s}}$ & $5.4\,g g_D$\\\T\B
$D^*\bar{D}^*\to \rho\rho$&$\sqrt{3}\,g g_D$ &$-\frac{\sqrt{3} g g_D (2 M^2_{D^*}+2 M^2_\rho-3s)}{2\,M^2_{D^*_s}}$ & $10.1\,g g_D$\\\T\B
$D^*\bar{D}^*\to \omega\omega$& $- g g_D$& $\frac{g g_D(2M^2_{D^*}+2M^2_\omega-3s)}{2\,M^2_{D^*}}$& $-5.8\,g g_D$\\\T\B
$D^*\bar{D}^*\to \phi\phi$&$0$ & $0$& $0$\\\T\B
$D^*\bar{D}^*\to J/\psi J/\psi$&$-2\,g_{D} g_{\eta_c}$ & $\frac{g_{D} g_{\eta_c}(2M^2_{D^*}+2M^2_{J/\psi}-3s)}{M^2_{D^*}}$& $-7.2\,g_{D} g_{\eta_c}$\\\T\B
$D^*\bar{D}^*\to \omega J/\psi$&$2\,g g_D$ &$-\frac{g g_D(2M^2_{D^*}+M^2_{J/\psi}+M^2_\omega-3s)}{M^2_{D^*}}$ & $9.5\,g g_D$\\\T\B
$D^*\bar{D}^*\to \phi J/\psi$& $0$&$0$ &$0$ \\\T\B
$D^*\bar{D}^*\to \omega\phi$& $0$&$0$ &$0$ \\\T\B
$D_s^*\bar{D}_s^*\to D_s^*\bar{D}_s^*$& $-2\,g^2_{D_s}$& $\frac{g^2_{D_s}(M^2_{J/\psi}+M^2_\phi) (4M^2_{D^*_s}-3s)}{2\,M^2_{J/\psi}M^2_\phi}$& $-18.3\,g_{D_s}^2$\\\T\B
$D_s^*\bar{D}_s^*\to K^*\bar{K}^*$& $\sqrt{2}\,g g_{D_s}$& $-\frac{g g_{D_s} (2M^2_{D^*_s}+2M^2_{K^*}-3s)}{\sqrt{2}M^2_{D^*}}$& $8.\,g g_{D_s}$\\\T\B
$D_s^*\bar{D}_s^*\to \rho\rho$&$0$ &$0$ &$0$ \\\T\B
$D_s^*\bar{D}_s^*\to \omega\omega$&$0$ &$0$ &$0$ \\\T\B
$D_s^*\bar{D}_s^*\to \phi\phi$&$-\sqrt{2}\,g g_{D_s}$ & $\frac{g g_{D_s}(2M^2_{D^*_s}+2M^2_{\phi}-3s)}{\sqrt{2}M^2_{D^*_s}}$& $-7.3\,g g_{D_s}$\\\T\B
$D_s^*\bar{D}_s^*\to J/\psi J/\psi$& $-\sqrt{2}\,g_{D_s} g_{\eta_c}$& $\frac{g_{D_s} g_{\eta_c}(2M^2_{D^*_s}+2M^2_{J/\psi}-3s)}{\sqrt{2}M^2_{D^*_s}}$& $-4.6\,g_{D_s} g_{\eta_c}$\\
\T\B$D_s^*\bar{D}_s^*\to \omega J/\psi$& $0$& $0$&$0$ \\\T\B
$D_s^*\bar{D}_s^*\to \phi J/\psi$&$2\,g g_{D_s}$ &$-\frac{g g_{D_s}(2M^2_{D^*_s}+M^2_{J/\psi}+M^2_\phi-3s)}{M^2_{D^*_s}}$ &$8.5\,g g_{D_s}$ \\\T\B
$D_s^*\bar{D}_s^*\to \omega\phi$&$0$ & $0$&$0$ \\
\end{tabular}
\caption{Amplitude projected in isospin and spin for $I=0$ and $J=2$. The approximate Total is obtained at the threshold of $D^*\bar{D}^*$.}
\label{tab:Am3}
\end{table}
\begin{table}[htb!]
\begin{tabular}{c|c|c|c}
 &Contact & V-exchange & $\sim Total[I[J^P]]$\\
\hline\hline
\T\B$D^*\bar{D}^*\to D^*\bar{D}^*$& $2\,g_D^2$ & $\frac{g_D^2\,(2 M_\omega^2 M_\rho^2+M_{J/\psi}^2 (- M_\omega^2+M_\rho^2)) (4 M_{D^*}^2-3 s)}{4\, M_{J/\psi}^2 M_\omega^2 M_\rho^2} $ & $0.6\,g^2_D$\\\T\B
$D^*\bar{D}^*\to K^*\bar{K}^*$& $-g g_D$& $-\frac{g g_D(2M_{D^*}^2+2M_{K^*}^2-3 s)}{4\,M^2_{D_s^*}}$& $1.2\,g g_D$\\\T\B
$D^*\bar{D}^*\to \rho\rho$&$0$ &$0$ &$0$ \\\T\B
$D^*\bar{D}^*\to \rho\omega$&$-2\sqrt{2}\,g g_D$ &$-\frac{g g_D(2M^2_{D^*}+M_\omega^2+M_\rho^2-3s)}{\sqrt{2}\,M^2_{D^*}}$ & $4.\,g g_D$\\\T\B
$D^*\bar{D}^*\to \rho J/\psi$& $4\,g g_D$& $\frac{g g_D(2M^2_{D^*}+M^2_{J/\psi}+M^2_\rho-3s)}{M^2_{D^*}}$& $-3.5\,g g_D$\\\T\B
$D^*\bar{D}^*\to \rho\phi$&$0$ & $0$& $0$\\
\end{tabular}
\caption{Amplitude projected in isospin and spin for $I=1$ and $J=0$. The approximate Total is obtained at the threshold of $D^*\bar{D}^*$.}
\label{tab:Am4}
\end{table}
\begin{table}[htb!]
\begin{tabular}{c|c|c|c}
 &Contact & V-exchange & $\sim Total[I[J^P]]$\\
\hline\hline
\T\B$D^*\bar{D}^*\to D^*\bar{D}^*$& $3\,g_D^2$ & $\frac{g_D^2\,(2 M_\omega^2 M_\rho^2+M_{J/\psi}^2 (- M_\omega^2+M_\rho^2)) (4 M_{D^*}^2-3 s)}{4\, M_{J/\psi}^2 M_\omega^2 M_\rho^2} $ & $1.6\,g^2_D$\\\T\B
$D^*\bar{D}^*\to K^*\bar{K}^*$& $\frac{9\,g g_D}{2}$& $\frac{g g_D(2M_{D^*}^4+M_{D^*}^2(4M^2_{D^*_s}+2M^2_{K^*}-3 s)+2M^2_{D^*_s}(2M^2_\rho-3s))}{4\,M^2_{D^*}M^2_{D_s^*}}$& $-2.5\,g g_D$\\\T\B
$D^*\bar{D}^*\to \rho\rho$&$\frac{3\,g g_D}{\sqrt{2}}$ &$\frac{g g_D(2M^2_{D^*}+2M_\rho^2-3s)}{2\sqrt{2}\,M^2_{D^*}}$ & $-1.3\,g g_D$\\\T\B
$D^*\bar{D}^*\to \rho\omega$&$0$ &$0$ & \\\T\B
$D^*\bar{D}^*\to \rho J/\psi$& $0$& $0$& $0$\\\T\B
$D^*\bar{D}^*\to \rho\phi$&$0$ & $0$& $0$\\
\end{tabular}
\caption{Amplitude projected in isospin and spin for $I=1$ and $J=1$. The approximate Total is obtained at the threshold of $D^*\bar{D}^*$.}
\label{tab:Am5}
\end{table}
\begin{table}[htb!]
\begin{tabular}{c|c|c|c}
 &Contact & V-exchange & $\sim Total[I[J^P]]$\\
\hline\hline
\T\B$D^*\bar{D}^*\to D^*\bar{D}^*$& $-g_D^2$ & $\frac{g_D^2\,(2 M_\omega^2 M_\rho^2+M_{J/\psi}^2 (- M_\omega^2+M_\rho^2)) (4 M_{D^*}^2-3 s)}{4\, M_{J/\psi}^2 M_\omega^2 M_\rho^2} $ & $-2.4\,g^2_D$\\\T\B
$D^*\bar{D}^*\to K^*\bar{K}^*$& $\frac{g g_D}{2}$& $-\frac{g g_D(2M_{D^*}^2+2M_{K^*}^2-3 s)}{4\,M^2_{D_s^*}}$& $2.7\,g g_D$\\\T\B
$D^*\bar{D}^*\to \rho\rho$&$0$ &$0$ &$0$ \\\T\B
$D^*\bar{D}^*\to \rho\omega$&$\sqrt{2}\,g g_D$ &$-\frac{g g_D(2M^2_{D^*}+M_\omega^2+M_\rho^2-3s)}{\sqrt{2}\,M^2_{D^*}}$ & $8.3\,g g_D$\\\T\B
$D^*\bar{D}^*\to \rho J/\psi$& $-2\,g g_D$& $\frac{g g_D(2M^2_{D^*}+M^2_{J/\psi}+M^2_\rho-3s)}{M^2_{D^*}}$& $-9.5\,g g_D$\\\T\B
$D^*\bar{D}^*\to \rho\phi$&$0$ & $0$& $0$\\
\end{tabular}
\caption{Amplitude projected in isospin and spin for $I=1$ and $J=2$. The approximate Total is obtained at the threshold of $D^*\bar{D}^*$.}
\label{tab:Am6}
\end{table}

\end{document}